\documentclass[a4paper,10pt]{edpsci} 
\usepackage{amsmath}
\usepackage{graphicx}
\usepackage{epstopdf}
\usepackage[numbers,sort&compress]{natbib}
\usepackage{hyperref}
\usepackage{comment}
\usepackage{url}
\usepackage{upgreek}
\usepackage{siunitx}

\usepackage{tabularx}
\usepackage{booktabs}
\usepackage{makecell}
\usepackage{multirow}

\usepackage[switch]{lineno}

\usepackage[T2A]{fontenc}

\hypersetup{colorlinks=true,citecolor=blue,urlcolor=blue,linkcolor=blue}

\usepackage[ruled]{algorithm2e}

\journalname{Journal of the European Optical Society-Rapid Publications (JEOS-RP)}

\articletype{Research article}

\begin{document}


\title{Benchmarking Dual-Polarization Silicon Nitride Photonic Integrated Circuits for Trapped-Ion Quantum Technologies}

\titlerunning{Benchmarking Dual-Polarization Silicon Nitride Photonic Integrated Circuits}



\author{Carl-Frederik Grimpe \inst{1}
\and
Anastasiia Lüßmann-Sorokina \inst{1,2,3}
\and
Guochun Du \inst{1}\correspondingauthor{\email{guochun.du@ptb.de}}
\and
Pragya Sah \inst{4,5}
\and
Steffen Sauer \inst{1,2,3}
\and
Elena Jordan \inst{1}
\and 
Rijil Thomas \inst{4}
\and
Pascal Gehrmann\inst{2,3}
\and
Maksim Lipkin \inst{4,5}
\and 
Stephan Suckow \inst{4}
\and
Max C. Lemme \inst{4,5}
\and 
Stefanie Kroker \inst{1,2,3}
\and 
Tanja E. Mehlstäubler \inst{1,6,7}
}




\institute{Physikalisch-Technische Bundesanstalt, Bundesallee 100, Braunschweig, 38116, Germany 
\and
        Technische Universität Braunschweig, Institute of Semiconductor Technology, Hans-Sommer-Str. 66, Braunschweig, 38106, Germany 
\and
           Laboratory for Emerging Nanometrology (LENA), Langer Kamp 6a/b, Braunschweig, 38106, Germany 
\and
            AMO GmbH, Otto-Blumenthal-Straße 25, 52074 Aachen, Germany
\and
        Chair of Electronic Devices, RWTH Aachen University, Otto-Blumenthal Str. 25, Aachen 52074, Germany
\and
           Leibniz Universität Hannover, Institut für Quantenoptik, Welfengarten 1, Hannover, 30167, Germany 
\and
           Leibniz Universität Hannover, Laboratorium für Nano- und Quantenengineering, Welfengarten 1, Hannover, 30167, Germany
           }

\abstract{Trapped ions are one of the most advanced platforms for quantum technologies, with applications ranging from quantum computing to precision timekeeping. A crucial step towards more compact and scalable systems involves integrating photonic integrated circuits (PICs) into surface ion traps to enable on-chip light delivery and optical addressing of individual ions. Currently, most implementations rely solely on transverse-electric (TE) mode grating couplers, where the emitted light is polarized in the plane of the chip. In this work, we design, fabricate and  characterize key silicon nitride (Si\(_3\)N\(_4\))  PIC components required for scaling trapped-ion based quantum systems to multiple operating zones, including incoupling structures, splitters, and grating couplers that support both TE and transverse-magnetic (TM) modes with comparable optical losses. We benchmark the PIC at 760\,nm, which is a typical wavelength for Yb$^{+}$-applications. The fabricated grating couplers enable the outcoupling of collimated free-space beams for both polarizations, exhibiting distinct emission angles. This dual-polarization capability gives more flexibility in polarization control and expands the accessible optical design space for trapped-ion quantum technologies.}





\maketitle

\section{Introduction}

Trapped ions are a leading platform for numerous quantum applications, including sensing \cite{biercuk_ultrasensitive_2010,Harlander_2010}, communications \cite{maunz2007quantum,higginbottom2016pure}, timekeeping \cite{Hausser25,Lindvall:25}, and computing \cite{Monroe13,Race_track23}.

The integration of nanophotonics into surface ion trap platforms has emerged as a promising route toward scalable quantum computing architectures and quantum sensors \cite{Mehta2016,Mehta2020,niffenegger2020integrated,Bruzewicz2019,Ivory21,Kwon2024,Knollmann2024,Mordini25}. Monolithically integrated waveguides (WGs) and grating couplers (GCs) enable stable, well-defined optical beam delivery with a small footprint, mitigating the alignment sensitivity of free-space optical systems. Importantly, these photonic components are compatible with CMOS processes, enabling wafer-scale fabrication \cite{Zesa2024}.


Among available material platforms, silicon nitride (Si\(_3\)N\(_4\)) has been developed as a mature and versatile option for visible and near-infrared operation \cite{Blumenthal2018,Munoz2017,Sacher2019,Leinse2020,Chauhan22,Smith23,Blasco-Solvas:24}. Si\(_3\)N\(_4\) offers low propagation loss and high optical-power handling \cite{Blumenthal2018}, together with CMOS-compatible fabrication suitable for both passive and nonlinear photonic devices. Advances in film-stress control, etch uniformity, and wafer-scale processing have improved reproducibility and yield \cite{Munoz2017,Leinse2020,10527127}, reinforcing Si\(_3\)N\(_4\) as a leading dielectric platform for visible integrated quantum technologies. 

To date, most Si$_3$N$_4$ photonic implementations in ion traps are optimized for transverse-electric (TE) polarization \cite{Mehta2016,Mehta2020,niffenegger2020integrated,Kwon2024,Mordini25}, where the electric field points parallel to the chip surface. The polarization of light determines the coupling to atomic transitions. Therefore, TE-optimized operation provides limited coupling directions, and placement options of the GC to deliver the correct polarization with respect to the ions quantization
axis. Furthermore, TE-optimized operation limits accessible beam geometries.
More recently, implementing transverse-magnetic (TM) modes, with the electric field oriented out of plane, have seen increased attention in integrated quantum and precision systems, such as Si$_3$N$_4$ resonator platforms for laser stabilization\,\cite{Isichenko2025}, and ion trap applications leveraging GCs for polarization gradient cooling\,\cite{corsetti_integrated-photonics-based_2026}. 


\begin{figure*}[ht]
\centering
\includegraphics[width=\textwidth,clip]{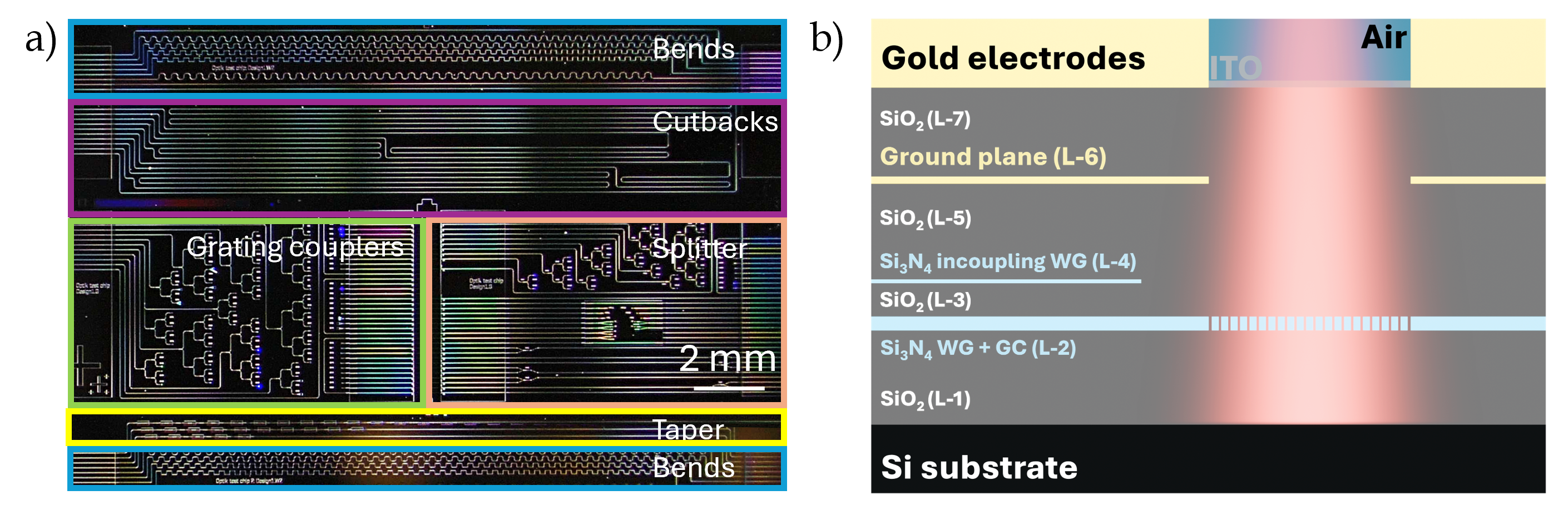}
\caption{ 
 a) Optical micrograph of the fabricated Si\(_3\)N\(_4\) PIC, showing different functional regions: WG cutback structures (purple), bend and bend-shift loss structures (cyan), EBL–OL taper structures (yellow), grating coupler (green) and splitter test structures (orange)).
b) Targeted multilayer stack. From bottom to top: Si substrate, SiO\(_2\) BOX (L-1), Si\(_3\)N\(_4\) WG-GC (L-2) and WG for light incoupling layers (L-4) separated by a 1.25\,$\upmu$m thick SiO\(_2\) buffer layer (L-3). 
Above these, an additional SiO\(_2\) buffer layer is followed by a gold ground plane, 3\,$\upmu$m thick SiO\(_2\) spacer, and a top gold layer used as the electrodes. Holes in the electrodes are covered with an ITO layer. The red cone illustrates the beam emitted by the GC.}
\label{fig-layer-stack}
\end{figure*}

We report the design, fabrication, and experimental characterization of the performance of a Si$_3$N$_4$ photonic integrated circuit (PIC) consisting of components such as incoupling structures, splitters, bends, taper sections, and GCs, that are key components for scaling trapped-ion based quantum technologies such as quantum computers and clocks  to multiple operating zones.  As an example, we use a wavelength of 760\,nm for applications with trapped  Yb$^{+}$ ions \cite{Huntemann2012,Tofful_2024}. The Yb$^{+}$ system underpins a wide range of state-of-the-art applications, including precision optical clocks \cite{Huntemann2012,Tofful_2024}, trapped-ion quantum information processing \cite{Olmschenk2007,piltz_trapped-ion-based_2014}, and tests of fundamental physics \cite{dreissen_improved_2022,Hur22,Filzinger25,Door25}. We characterize the PIC with TE and TM modes and discuss how the observed polarization dependence may be leveraged to extend the photonic toolbox for ion trap quantum technology applications by offering additional design flexibility and enabling dual-polarization architectures.

\section{TE \& TM light guiding}
\label{sec-light}

\subsection{Chip design and fabrication}

The chip layout (Fig.\,\ref{fig-layer-stack}\,a)) comprises several functional sections, including WG cutback structures for the characterization of propagation losses, bend and bend-shift test structures, taper test structures, GC studies, and splitter test structures designed to operate at 760\,nm. Together, these sections enable a comprehensive evaluation of polarization dependence, coupling efficiency, and loss mechanisms across the PIC. Each section is highlighted in a different color, corresponding to distinct device groups used to characterize optical performance and assess design variations.

The targeted multilayer trap stack comprises both dielectric and metallic layers, as depicted in Fig.\,\ref{fig-layer-stack}\,b). Starting from the silicon (Si) substrate, a thermally grown 3\,$\upmu$m thick buried oxide (BOX)  silicon dioxide (SiO\(_2\)) layer acts as the lower cladding (L-1). The thickness of this bottom oxide layer is chosen to suppress optical leakage into the silicon substrate. On top of this, a 200\,nm thick low-pressure chemical vapor deposition (LPCVD) Si\(_3\)N\(_4\) layer (L-2) is deposited. An intermediate 1.25\,$\upmu$m thick low-temperature oxide (LTO) SiO\(_2\) spacer (L-3) separates this layer from a 15\,nm thick LPCVD Si\(_3\)N\(_4\) layer (L-4). Owing to its larger optical mode, this thin Si\(_3\)N\(_4\) layer is positioned at the top of the stack to maximize the separation from the silicon substrate and thereby minimize substrate leakage. An additional 3\,$\upmu$m thick LTO SiO\(_2\) layer (L-5) forms the upper cladding of the PIC and is designed to be sufficiently thick to prevent optical leakage of the guided modes into the metallic layers introduced above. 
The targeted stack is completed by a 100\,nm thick gold ground plane (L-6) introduced to provide electrical isolation and shielding, followed by a 3\,$\upmu$m thick SiO\(_2\) buffer layer (L-7). A top gold layer forming the ion trap electrodes completes the envisioned ion trap architecture, with a thin indium tin oxide (ITO) layer included as a transparent conductive material covering holes in the electrodes that could otherwise affect the trapped ion \cite{du2025electricfielddistortionssurface}.

For the fabrication of the PIC, a mix-and-match lithography approach was employed. Features larger than 500\,nm, such as WGs and splitters, were realized using optical lithography (OL), whereas smaller features, such as incoupling tapers, were patterned with electron-beam lithography (EBL). 
Consequently, the Si\(_3\)N\(_4\) GCs were fabricated in two sequential lithographic steps, the taper region defined by OL and the grating teeth subsequently patterned by EBL to achieve the required submicrometer periodicity and duty-cycle precision. The Si\(_3\)N\(_4\) layer thickness of 200\,nm was chosen as a compromise between compatibility with OL-based waveguide fabrication, GC coupling efficiency, and the minimum feature sizes achievable by EBL. \\
All photonic designs and simulations presented in this work are carried out with the full target layer stack in mind.
The experimental characterization reported here is performed on representative photonic test structures fabricated only up to layer L-5, before the deposition of the metallic layers.

\subsection{Photonic component design}
\subsubsection{Alignment-tolerant incoupling}

\begin{figure}[ht]
\centering
\includegraphics[width=8cm,clip]{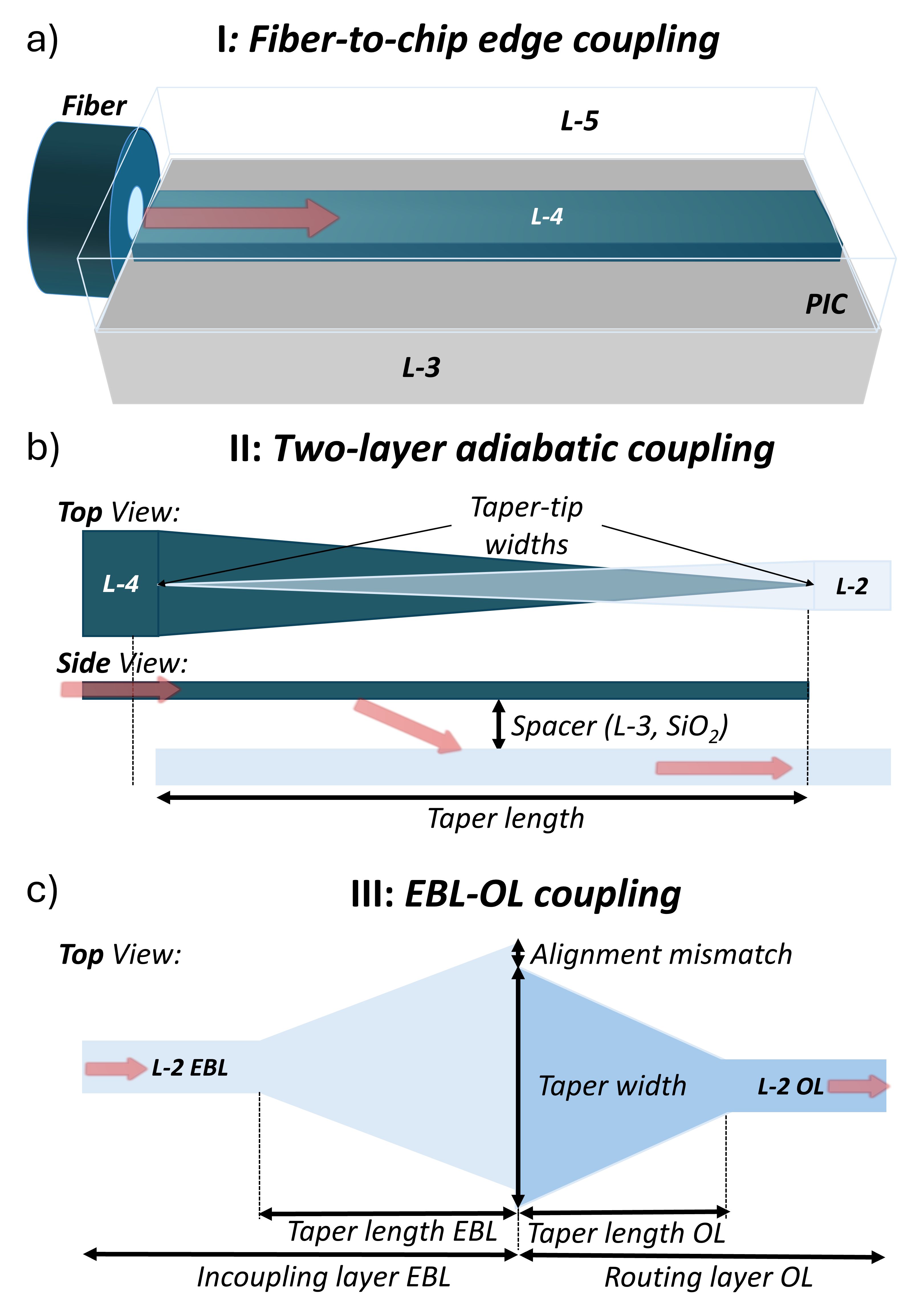}
\caption{Schematics of vertical fiber-to-chip coupling scheme. (a) Illustration of the fiber-to-chip edge coupling into the  PIC. (b) Geometry of the two-layer adiabatic taper for efficient mode conversion between L-4 and L-2 waveguides. (c) Schematic of the EBL-to-OL coupling stage. The red arrows depict the light propagation through the structures.}
\label{incoupling}
\end{figure}

The incoupling region consists of three functional parts employing two vertically separated Si$_3$N$_4$ layers.
The thinner upper layer serves for mode matching to the input fiber (see Fig.\,\ref{incoupling}\,a), providing an expanded optical mode comparable to the fiber mode field diameter, while the thick-core lower layer serves as the routing layer for on-chip propagation. 
Light is transferred adiabatically between the two layers by a vertically tapered section (see Fig.\,\ref{incoupling}\,b), which minimizes mode conversion loss through gradual effective index evolution \cite{Sun:09}. 
Such dual-layer and vertically adiabatic coupling concepts are widely adopted in Si$_3$N$_4$ and Si-based photonic platforms to achieve efficient fiber-to-chip coupling, sub-dB loss, and relaxed alignment tolerances\,\cite{Lin2021_BiLayerSiN,Zhang2021_Buried3DSSC,Dai2006_BilevelConverter,Jia2018_SuspendedCoupler,Yi2024_BilevelDualCore,Chen2025_DualSiOxN,He2020_JLT_EdgeCouplers,Marchetti2019_CouplingReview,Mu19}. 
Finally, a second taper region is introduced to compensate for potential lateral misalignment between the EBL and OL process levels, ensuring robust and efficient coupling (see Fig.\,\ref{incoupling}\,c)).

The designs were carried out using Ansys Lumerical, employing the Finite Difference Eigenmode (FDE) solver for mode-overlap simulations to optimize fiber-to-chip coupling and eigenmode expansion (EME) simulations were used to analyze adiabatic transitions and lithography overlap sections.~The simulated total incoupling loss is 1.95\,dB for TE and 1.65\,dB for TM, assuming perfect alignment across all fabrication steps. When considering a 500\,nm lateral misalignment at each coupling stage, the total incoupling loss increases to 2.26\,dB (TE) and 2.03\,dB (TM), demonstrating the alignment-tolerance of the incoupling design. ~Resulting design parameters, coupling efficiencies, and simulations are summarized in the Supplementary Information (SI).

\subsubsection{Routing}

To deliver light to the ions, it is routed across the chip via WGs, bends, and splitters before being directed into the GC.

\paragraph{\textit{Waveguides}}

WG modes were simulated using the FDE solver and the final waveguide geometry was then fixed to 200\,nm\,$\times$\,520\,nm (height\,$\times$\,width) to secure single-mode operation and avoid higher-order contributions in the GC emission (see SI).

\paragraph{\textit{Bends}}

Bend structures with radii of 10, 20, 40, and 60\,$\upmu$m were designed, along with bend-shift structures optimized for TE mode matching inbetween bent and straight WGs using the FDE solver \cite{Sakai_bendshift}. The principle is illustrated in the inset of Fig. \ref{bends} a). Bend shifts $\Lambda$ of 30, 10, 10 \,nm were implemented for the 10, 20, 40\,$\upmu$m bends, respectively. They offer a promising approach to lowering mode-mismatch losses in compact bends, which in turn increases the flexibility of the photonic design \cite{Sakai_bendshift}.

\paragraph{\textit{Splitter}}

multimode interference (MMI) splitters are common on-chip components used for optical power splitting based on self-imaging effects in multimode waveguides \cite{Soldano_MMI}. The MMI splitters were designed using EME simulations for TE-polarized light (see SI). Each splitter features a core length of 38\,$\upmu$m, a core width of 6\,$\upmu$m, taper sections of 10\,$\upmu$m length with 1.6\,$\upmu$m width, and an output channel separation of 3.14\,$\upmu$m.

\subsection{Experimental comparison of TE and TM losses}

\subsubsection{Setup}
A fiber-coupled 760\,nm distributed Bragg reflector (DBR) laser \footnote{Thorlabs DBR760PN} was used as the light source for the characterization of the PIC. The input polarization was adjusted, and coupling was achieved using a single-mode (SM) polarization-maintaining (PM) \footnote{PM780-HP} fiber  aligned with a nanopositioner \footnote{PI P-616 NanoCube® XYZ-Nanopositionierer}. A powermeter was used to measure the optical power, while a pick-off on a photodiode was measured to monitor laser power fluctuations during the measurements. The polarization state was set using a polarimeter\footnote{SK010PA-NIR} (see Fig.\,\ref{incoupling_meas}\,a)).

Experiments were performed with either a custom-made fiber array with PM fibers in a U-configuration, aligned to dedicated reference structures at the outer waveguide channels, or by coupling light through a single PM fiber of the array and collecting the output with a SM fiber\footnote{SM 780HP} positioned behind the chip using a three-axis alignment system \footnote{NanoTrak® Auto-Alignment Controller}.

Optical measurements and characterization of the GC emission were performed using a commercial microscope system \footnote{Olympus BXFM} equipped with a high-NA objective \footnote{Olympus MPLAPON50X} (0.95 NA) and a CMOS camera\footnote{Hamamatsu ORCA-spark C11440-36U}. The PIC is mounted on a tip, tilt and rotation stage \footnote{Thorlabs TTR001} to minimize tilt-induced angle errors.  In addition, the grating efficiency was determined by summing the pixel intensities of the outcoupled light on the CMOS camera and converting them to optical power using a pre-established calibration factor.
Prior to these measurements, a reference calibration was performed using a fiber-coupled source to evaluate the objective transmission and to determine the conversion coefficient relating pixel intensity to optical power per exposure time.

\subsubsection{Measurements}
\begin{figure}[h]
\centering
\includegraphics[width=8cm,clip]{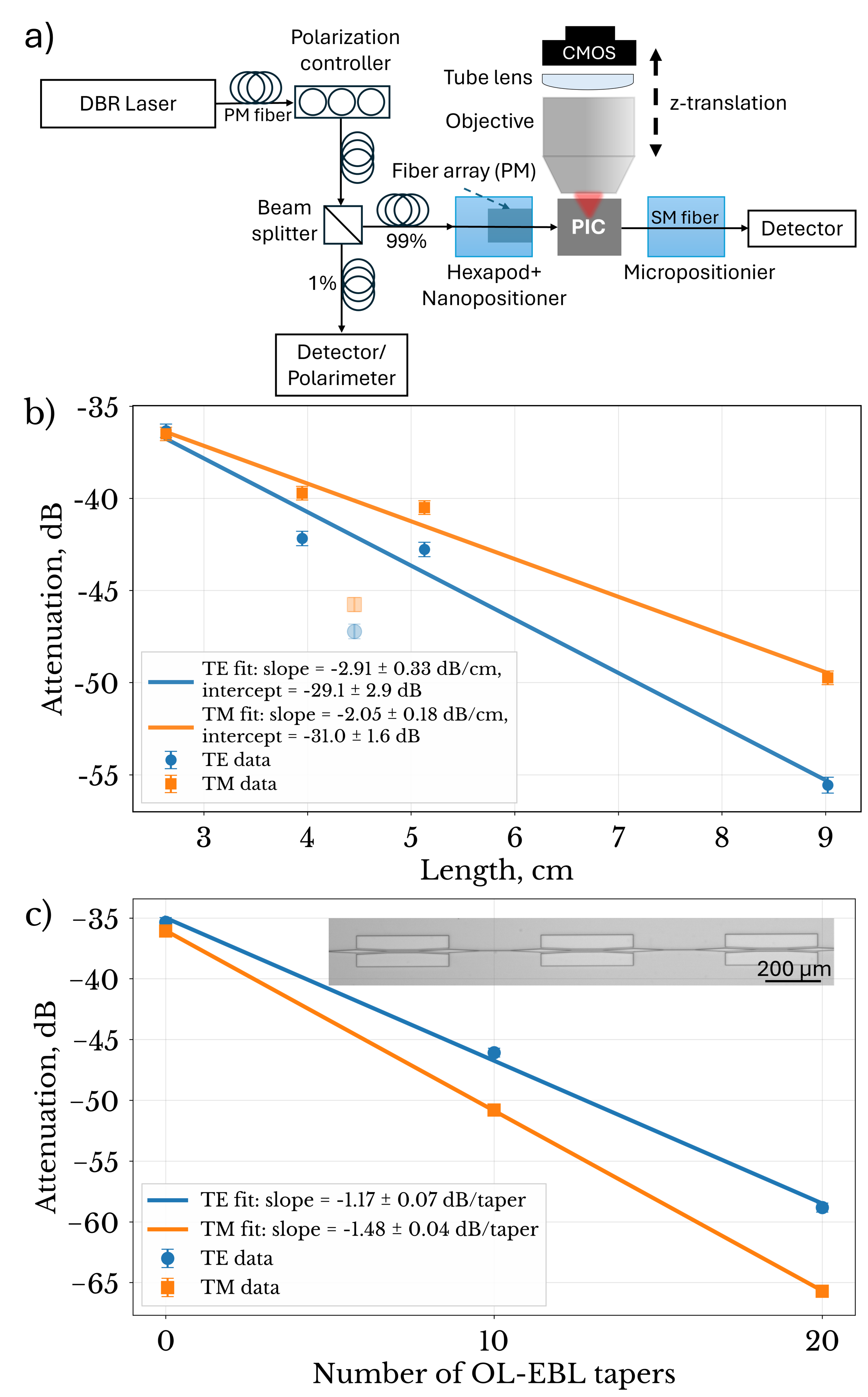}
\caption{a) Schematic illustration of the optical characterization setup. The red cone illustrates the beam emitted from the PIC.
b) Cutback measurement to deduce the coupling and propagation losses for the TE and TM modes at a wavelength of 760\,nm. Data points shown with reduced opacity were excluded from the fit.
c) Characterization of the EBL–OL transition taper, showing the extracted taper loss. The inset displays the cascaded taper section.}

\label{incoupling_meas}
\end{figure}

\paragraph{\textit{Propagation losses}}
Propagation losses were determined by measuring the transmitted power through waveguides of varying lengths arranged in a U-configuration. The exponential decay of transmitted power with length was fitted to extract the loss parameters, where the slope corresponds to the propagation loss and the intercept represents twice the combined incoupling and bend losses (see Fig.\,\ref{incoupling_meas}\,b)). Measured propagation losses are (2.91\,$\pm$\,0.33)\,dB/cm for the TE mode and (2.05\,$\pm$\,0.18)\,dB/cm for the TM mode.
\paragraph{\textit{Taper losses}}
Ol-EBL taper losses have been characterized to be (1.17\,$\pm$\,0.07)\,dB per taper for TE and (1.48\,$\pm$\,0.04)\,dB per taper for TM  by fitting the decay of the power transmission through waveguides with a different number of taper sections (see Fig.\,\ref{incoupling_meas}\,c). The OL-EBL taper losses are contributing to the total incoupling loss since they directly capture the loss introduced by the mix-and-match lithography approach.

\begin{figure*}[ht]
\centering
\includegraphics[width=\textwidth,clip]{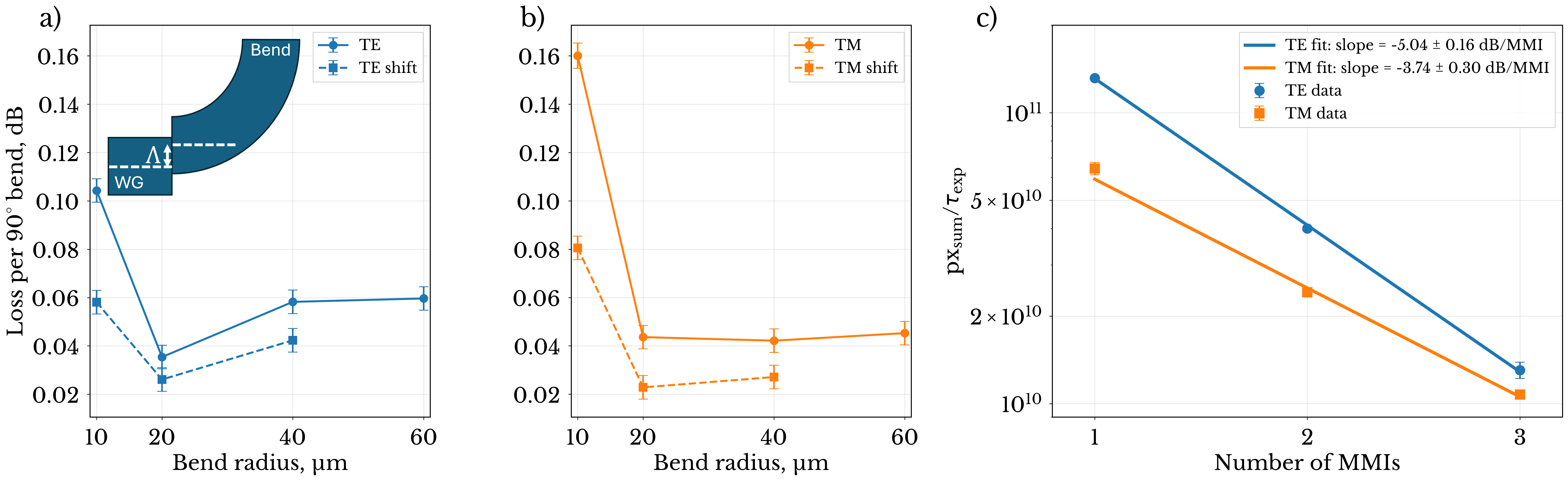}
\caption{
Per-bend loss for the TE a) and TM b) polarization, comparing standard bends to laterally shifted bend geometries across different bend radii. In the shifted designs, 
$\Lambda$ denotes the lateral offset, as illustrated in the inset of a). 
c) Measured attenuation of cascaded MMI splitters for the TE and TM modes. The linear fits provide the attenuation per MMI device and include additional contributions from bend and propagation losses.
}

\label{bends}
\end{figure*}
\paragraph{\textit{Bend and bend-shift losses}} 

Bend losses were determined by comparing waveguides containing bends with varying radii to a straight reference waveguide. The extracted values include the intrinsic propagation loss within the curved sections. The total excess loss was then divided by the number of bends (200) to obtain the bend loss per bend.

The corresponding bend-loss measurements are shown in Fig.\,\ref{bends}\,(a,b) for TE- and TM-polarized light, respectively. For both polarizations, the bend loss strongly depends on the bend radius and is significantly reduced by the bend-shift design. For TE polarization (Fig.\,\ref{bends}\,a)), bends with radii of 20, 40, and 60\,\(\upmu\)m exhibit losses below 0.06\,dB per 90\(^\circ\) bend. The lowest loss is observed for the 20\,\(\upmu\)m bend, with approximately 0.04\,dB, while the corresponding bend-shift design further reduces the loss to below 0.03\,dB. At larger radii, the loss remains nearly constant, indicating that radiation and mode mismatch losses are already well suppressed in this regime.
For TM polarization (Fig.\,\ref{bends}\,b)), a similar trend is observed. For bend radii of 20\,\(\upmu\)m and above, losses remain low, with values of approximately 0.04\,dB for a 40\,\(\upmu\)m bend and as low as 0.02\,dB for the 20\,\(\upmu\)m bend-shift design. As for TE polarization, the bend-shift geometry consistently outperforms the standard bend, confirming its effectiveness in reducing mode mismatch loss for both polarizations.
At a bend radius of 10\,\(\upmu\)m, a pronounced increase in loss is observed for both polarizations, exceeding 0.10\,dB per bend. In this regime, TE losses increase to approximately 0.11\,dB, while TM losses reach about 0.16\,dB, indicating the onset of radiation and mode mismatch losses due to insufficient mode confinement at tight bend radii.
\paragraph{\textit{MMI splitter}} 
The splitter losses were evaluated using an MMI splitter tree terminated by grating couplers, each configuration containing a different number of MMIs. The total transmitted optical power was obtained from the summed pixel intensity $\mathrm{px_{sum}}$ per exposure time $\tau_{\mathrm{exp}}$ monitored by the CMOS camera and plotted logarithmically as a function of the number of MMIs. The extracted slope represents the combined loss per MMI arm of one MMI, two bends, and a 150\,$\upmu$m straight waveguide section. From this, the individual MMI insertion loss was determined to be  (1.88\,$\pm$\,0.16)\,dB for TE and (0.62\,$\pm$\,0.30)\,dB for TM (see Fig.\,\ref{bends}\,c)). 
\paragraph{\textit{Incoupling} }
The incoupling loss was estimated in two ways: (i) from propagation-loss measurements by subtracting the contributions from bends and EBL–OL taper sections from the y-intercept, and (ii) from direct transmission measurements through a straight waveguide across the chip, with propagation losses subtracted. Method (i) yields single-sided incoupling losses of (13.0\,$\pm$\,1.5)\,dB for TE and (14.08\,$\pm$\,0.81)\,dB for TM, while method (ii) gives (13.89\,$\pm$\,0.43)\,dB for TE and (14.71\,$\pm$\,0.43)\,dB for TM (see Fig.\,\ref{incoupling_meas}\,b)). These incoupling loss values correspond to the sum of the fiber-to-chip edge-coupling loss and the loss introduced by the two-layer adiabatic coupler. The total incoupling loss per facet is given by these contributions, including additional EBL-OL taper loss.
The measured losses are significantly higher than those predicted by simulation, which may be attributed to fabrication-induced deviations in the taper-tip widths.
\section{Design and characterization of the grating couplers}
\label{sec-1}
\begin{figure*}[ht]
\centering
\includegraphics[width=\textwidth,clip]{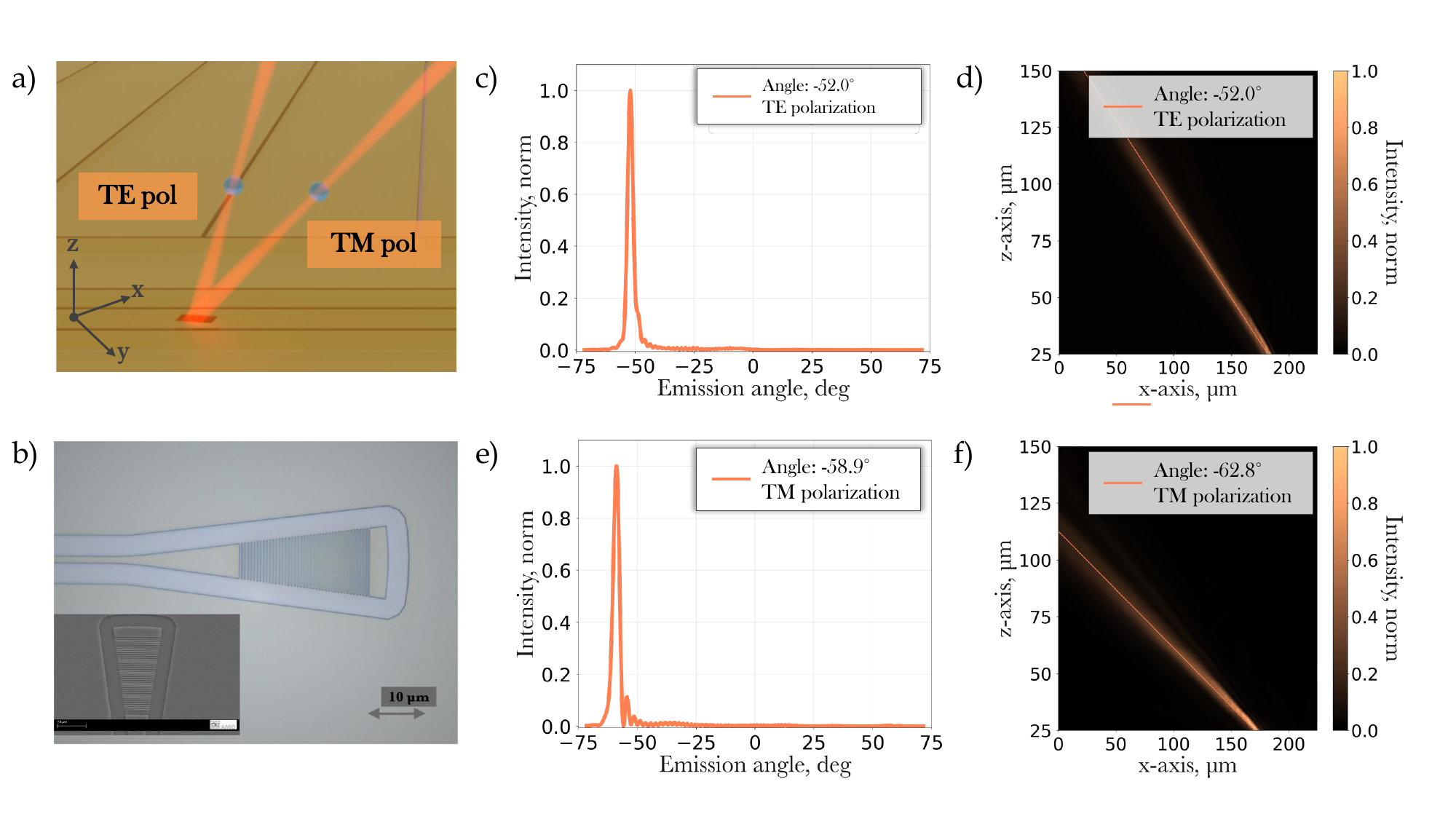}
\caption{
a)\,Schematic illustration of the two beams emitted from the GC, corresponding to orthogonal TE and TM polarizations. 
b)\,Top-view optical microscope image of the fabricated Si\(_3\)N\(_4\)/SiO\(_2\) GC.
(c,e)\,Simulated far-field intensity distributions as a function of emission angle, evaluated in a plane 100\,$\upmu$m above the chip surface, for TE and TM polarizations, respectively.  
(d,f)\,Experimental intensity maps acquired from 25\,$\upmu$m to 150\,$\upmu$m above the chip surface were fitted to reconstruct the emission profiles of the GC for TE and TM polarizations, respectively.}
\label{GC_angles}
\end{figure*}
The operational principle of the structure is illustrated schematically in Fig.\,\ref{GC_angles}\,a). 
Two beams corresponding to orthogonal TE and TM polarizations are emitted from a single GC and through an opening in the gold electrode layer.
An optical microscope image of the fabricated device is presented in Fig.\,\ref{GC_angles}\,b).
\paragraph{\textit{Design}}The design optimization was performed using a combination of 2D and 3D finite-difference time-domain (FDTD) simulations via Ansys Lumerical.
In the 2D model, the grating periods and duty cycles were varied to control the emission angle and far-field beam shape for the TE polarization \cite{Taillaert:04,Gill2024,2016Oton}. 
To finalize the design, a 3D FDTD simulation was used to define the grating taper geometry and grating taper width and to extract the far-field intensity profile from the resulting structure. The GC design is primarily optimized for TE operation, resulting in a 30\,$\upmu$m-long grating section for efficient TE outcoupling, while simultaneously supporting a well-defined TM-polarized response, enabling a comprehensive assessment of polarization selectivity, angular separation of the emitted beams, efficiency and their quality.

\paragraph{\textit{Measurement}}
The outcoupled intensity distribution was recorded slice by slice in the xy-plane by translating the microscope system along the z-direction with a step size of 0.5\,$\upmu$m. 
At each $z$-slice, the intensity profile of the emitted beam was analyzed by determining the position of maximum intensity. This maximum was taken as the beam position in the transverse plane. By repeating this procedure for all measured $z$-slices, we obtained a set of beam-center coordinates along the propagation direction. The emission angle was then extracted by performing a linear fit to these beam-center positions.
The resulting angles for both polarizations were then compared with simulations, as shown in Fig.\,\ref{GC_angles}\,(c–f). 
The measured emission angle of ($-52.0 \pm 0.4)^\circ$ for TE polarization agrees well with the simulated value of $-52.0^\circ$. For TM polarization, the measured angle of ($-62.8 \pm 0.4)^\circ$ shows a larger deviation from the simulated value of $-58.8^\circ$. This discrepancy may be attributed to the increased sensitivity of steeper emission angles to fabrication tolerances. The measurement uncertainty accounts for the individual angle fit error per measurement ($\pm0.1^\circ$ for both TE and TM), the uncertainty in the z-translation calibration of the microscope system ($\pm0.3^\circ$ for both TE and TM), and the chip-tilt uncertainty ($\pm0.2^\circ$).

Angles are reported with respect to the surface normal to ensure consistency between simulations and the experimental analysis. Negative angles correspond to a backward-emitting GC.

\begin{figure*}[!ht]
\centering
\includegraphics[width=\textwidth,clip]{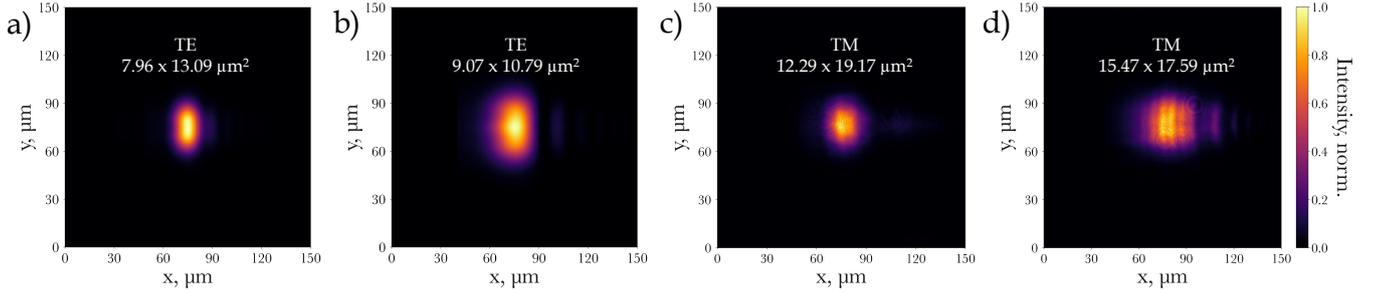}
\caption{
 Simulated (a, c) and measured (b, d) far-field intensity distributions of the laser-beam waist at 100\,$\upmu$m above the chip surface for TE (a, b) and TM polarization (c, d).}

\label{GC_beams}
\end{figure*}

A detailed comparison between the simulated and measured far-field beam profiles is presented in Fig.\,\ref{GC_beams}\,, where all data sets were analyzed using the elliptical contour method and 1D Gaussian fits along the principal axes. The ellipse-based method provides a geometric estimate of the beam diameters along the principal axes of the far-field spot, while the Gaussian model fitting yields beam waists with one-standard-deviation uncertainties.
The TE beam exhibits simulated waists of $(7.96\pm0.01)\,\upmu\mathrm{m}\times(13.09\pm0.01)\,\upmu\mathrm{m}$, compared to measured waists of $(9.07\pm0.24)\,\upmu\mathrm{m}\times(10.79\pm0.24)\,\upmu\mathrm{m}$. For the TM polarization, the simulated beam waists are $(12.29\pm0.01)\,\upmu\mathrm{m}\times(19.17\pm0.01)\,\upmu\mathrm{m}$, while the measured beam waists are $(15.47\pm0.24)\,\upmu\mathrm{m}\times(17.59\pm0.24)\,\upmu\mathrm{m}$.

The simulated grating-coupler efficiencies are \(-3.61\)\,dB for TE polarization and \(-8.79\)\,dB for TM polarization. These values are in good agreement with the experimentally measured efficiencies of \((-3.55 \pm 0.58)\)\,dB for TE and \((-9.02 \pm 0.69)\)\,dB for TM, which were obtained by measuring the emitted free-space power relative to the power after the fiber array and subtracting the losses of the remaining PIC components.

\section{Discussion}

\paragraph{\textit{Propagation losses} }Both TE- and TM-polarized light were guided in the fabricated Si$_3$N$_4$ waveguides. The measured propagation loss is $0.86 \pm 0.38\,\mathrm{dB/cm}$ higher for the TE polarization than for the TM polarization. This behavior indicates that the propagation loss is dominated by sidewall roughness, as the TE mode exhibits a larger modal overlap with etched sidewalls. The measured losses remain higher than state-of-the-art Si$_3$N$_4$ waveguides of similar thickness, where values as low as 2\,dB at 729\,nm have been demonstrated \cite{Mehta2020}. This indicates that further optimization of the fabrication process is required. The TM propagation loss may be attributed to leakage into the ground plane or substrate, since their larger vertical mode extent demands sufficiently thick oxide layers to suppress. Transitioning to fused silica substrates would further mitigate substrate-related losses \cite{Dietl25}.
\paragraph{\textit{Incoupling} }Incoupling losses of around 14\,dB exceed the simulated values, yet both evaluation methods yielded similar absolute coupling efficiencies, with TE coupling slightly lower than TM. 
For comparison, visible-wavelength dual-layer Si$_3$N$_4$ couplers have been reported with losses of about 4\,dB/facet \cite{Lin2021_BiLayerSiN}, while single-layer couplers typically reach around 7.5--8.5\,dB/facet \cite{Sacher2019}. The values obtained here are higher than those reported, suggesting that additional loss mechanisms are present in the fabricated devices. This behavior is plausibly linked to the two-layer edge-coupler architecture, which is highly sensitive to the taper-tip width of the thick-core Si$_3$N$_4$ layer \cite{Lin2021_BiLayerSiN}. A taper-tip wider than the optimized target value reduces the adiabaticity of the local index transition near the taper entrance and diminishes modal overlap (see SI). This indicates that future designs require an improved fabrication of the taper tip, or a thinner thick-core Si$_3$N$_4$ layer to reduce sensitivity to the taper-tip width. Measurements with the fiber array yielded coupling efficiencies comparable to the single-fiber configuration, indicating that alignment constraints introduce no significant additional penalty.
\paragraph{\textit{Bend and bend-shift losses} } The measured bend losses of around 0.04 \,dB per 20-$\upmu$m-radius bend match results reported in visible-wavelength Si$_3$N$_4$ experiments \cite{Sacher2019}. TM-polarized light exhibited noticeably higher loss for small-radius bends, a consequence of the 200\,nm Si$_3$N$_4$ thickness, which was optimized for TE operation and provides insufficient confinement for TM, thereby increasing curvature-induced radiation loss. Such polarization-dependent bend performance aligns with earlier analyses of thin-core Si$_3$N$_4$ waveguides \cite{Munoz2017}. The shift of the bend-loss minima to larger radii —from 20\,$\upmu$m for TE to 40\,$\upmu$m for TM modes— is consistent with the lower propagation loss observed for TM.

\begin{table*}[!ht]
\centering
\caption{Comparison between simulated and measured losses for different PIC components for TE- and TM-polarized modes at 760\,nm.}
\label{tab:loss_comparison}
\begin{tabular}{lcccc}
\hline
\textbf{Component} & \multicolumn{2}{c}{\hspace{-0.9cm}\textbf{TE loss}} & \multicolumn{2}{c}{\hspace{-0.9cm}\textbf{TM loss}} \\
 & Simulated & Measured & Simulated & Measured \\
\hline
Waveguide & & \hspace{0.6cm}$2.91  \pm 0.33\ \mathrm{dB/cm}$& & \hspace{0.6cm}$2.05 \pm 0.18\ \mathrm{dB/cm}$ \\
Incoupling & $1.95\ \mathrm{dB}$ & \hspace{-0.25cm} $13.89\pm 0.43\ \mathrm{dB}$ & $1.65\ \mathrm{dB}$ & \hspace{-0.25cm} $14.71 \pm 0.43\ \mathrm{dB}$ \\
EBL-OL taper & $1.09\ \mathrm{dB}$ & $1.17 \pm 0.07\ \mathrm{dB}$ & $0.92 \ \mathrm{dB}$ & $1.48 \pm 0.04\ \mathrm{dB}$ \\
MMI splitter & $0.77\ \mathrm{dB}$ & $1.88 \pm 0.16 \ \mathrm{dB}$  & $0.42\ \mathrm{dB}$ & $0.62 \pm 0.30 \ \mathrm{dB}$ \\
GC & $3.61\ \mathrm{dB}$ & $3.55 \pm 0.58\ \mathrm{dB}$ & $8.79\ \mathrm{dB}$ & $9.02 \pm 0.69\ \mathrm{dB}$ \\
\hline
\end{tabular}
\end{table*}

Introducing bend-shift structures proved particularly beneficial for tight bends, where they reduced excess loss associated with mode mismatch at straight-to-bend transitions. This strategy is valuable for compact photonic-circuit layouts or applications requiring minimal stray light and crosstalk, as it enables dense routing without compromising optical efficiency.
\paragraph{\textit{MMI splitter} }Significant deviations from simulated values were observed for TE MMI splitter insertion losses, while TM performance remained close to predictions. Enhanced TE losses can be attributed to stronger sensitivity to sidewall scattering within the multimode region, which perturbs the self-imaging condition. This observation is consistent with the higher TE propagation losses across the platform. TM modes, with weaker confinement, are less susceptible to such imperfections. Further optimization of the MMI length would likely reduce the residual insertion loss (see SI).
\paragraph{\textit{Taper losses}} The OL-EBL taper transitions show good agreement with simulations for the TE polarization, while larger deviations are observed for the TM mode. These deviations may arise from lateral and longitudinal misalignment, as well as propagation losses, which are not accounted for in the simulations. Additionally, the observed behavior can be attributed to polarization-dependent changes in the modal field distribution induced by the taper. For the TE mode, tapering reduces the overlap with the etched sidewalls, thereby lowering sidewall-scattering loss. In contrast, for the TM mode, tapering may enhance propagation loss by increasing the overlap of the modal field with horizontal layer interfaces, thereby amplifying interface-related loss mechanisms. Extending the taper length or employing higher-order polynomial or spline profiles could further improve adiabaticity and reduce insertion loss. 
\paragraph{\textit{Grating coupler}}The measured grating emission profiles closely match the simulated far-field profiles in both overall shape and divergence. When quantified using Gaussian beam waists along the fitted principal axes, the average absolute deviation between simulated and measured data is 17-18\,\% for both polarization regimes. The TM profiles exhibit a side lobe originating from the substrate back-reflected beam, which becomes visible because the lateral offset between the upward- and downward-reflected beams increases at larger angles. 
The measured emission angles are likewise consistent with the simulated values: for TE polarization, the emission angle matches the simulated value and agrees within experimental uncertainty, whereas for TM polarization, a deviation of approximately 4\(^\circ\) is observed. 
These differences may be attributed to fabrication tolerances, such as slight variations in the etch depth of the grating and the duty cycle of the non-uniform grating design. They may also be attributed to deviations from the targeted layer stack, which arose from the omission of the upper metallic and dielectric layers in the devices that were characterized experimentally.
The more pronounced deviation observed for the TM mode is consistent with the increased sensitivity of grating couplers emitting at larger angles, where small angular variations within the SiO\(_2\) cladding translate into larger deviations in free space.
Additional contributions may arise from the finite numerical aperture of the collection optics, minor phase-front distortions, and aberrations in the imaging system.
Overall, the beam morphology, emission direction, and polarization-dependent divergence remain in good agreement, confirming that the fabricated grating accurately reproduces the simulated optical performance.\\
The GC efficiencies also align well with simulated predictions (see Tab.\,\ref{tab:loss_comparison}). Small deviations may originate from small fabrication-induced variations. The lower coupling efficiency observed for the TM grating coupler is attributed to its weaker grating coupling strength. This behavior of TM GCs may be beneficial in the design of highly focusing GCs or GCs that emit large beams, where precise control over the outcoupling strength is essential. The observed 8–10$^\circ$ angular separation between the TE and TM beams enables two distinct operation regimes: (i) individual addressing of ions or emitters at spatially separated locations when the beam waist is small, or (ii) polarization-state discrimination by monitoring the relative output intensity at the TE- and TM-specific emission angles. The angular separation corresponds to a spacing of roughly 14–18\,$\upmu$m at an ion height of 100\,$\upmu$m. This separation can be tuned through the waveguide geometry and grating duty cycle, which set the offset in effective index between the modes and thus control the resulting beam spacing. Accordingly, this effect is more pronounced in high-index platforms. This effect is fundamentally limited by the requirement that both modes satisfy the Bragg condition; consequently, at large backward emission angles, the other polarization may not yet be present.

\section{Conclusion}
\label{sec-con}

We designed key  Si$_3$N$_4$  PIC components and experimentally characterized their polarization-dependent performance   for ion trap applications. The fabricated GCs support outcoupling for both polarizations and exhibit distinct emission angles. \\ This dual-polarization capability expands the accessible optical design space and may be used for multi-ion addressing through angular multiplexing, enabling more compact and versatile on-chip architectures while reducing the number of required GCs in integrated ion trap platforms. At the same time, the ability to excite both modes highlights the importance of robust polarization management, as inadvertent TM excitation can generate parasitic beams and increase stray light in the trapping region. Robust solutions such as TE–TM splitters or polarization converters can help to ensure a well-defined input state \cite{Hattori:24,Gallacher22}. The weaker coupling strength of the TM mode is promising for the design of highly focusing GCs for single-ion addressing in trapped-ion quantum processors or GCs that emit large beams for multi-ion addressing in multi-ion clocks.\\
Furthermore, since TM modes exhibit stronger sensitivity to substrate leakage, transitioning to fused-silica substrates \cite{Dietl25} or employing thicker SiO$_2$ BOX cladding layers may be advantageous in future implementations. Taken together, TE–TM operation introduces a useful form of polarization diversity that complements spatial-multiplexing strategies explored in multimode photonics \cite{Momenzadeh2025} and provides a promising pathway toward densely integrated, flexible, and scalable photonic interfaces for next-generation multi-zone trapped-ion quantum technologies.

\acknowtext

 We thank Markus Kromrey for taking images of the chips and Fatemeh Salahshoori for preparation of the chips.

\funding

The authors gratefully acknowledge the support by the Quantum Valley Lower Saxony, the BMFTR-project ATIQ (FKZ: 13N16116, 13N16126, 13N16130), Braunschweig International Graduate School of Metrology B-IGSM, the cluster of Excellence Quantum Frontiers (EXC-2123-390837967), and the Cluster of Excellence PhoenixD (EXC-2122).

\conflict
The authors declare that they have no competing interests to report.

\dataavailability

Data supporting the findings of this study are available from the corresponding author upon reasonable request.

\authorcontrib

Conceptualization, C.-F.G., G.D., S.Sa., A.L-S., E.J., T.M., S.K., S.Su.; Methodology, C.-F.G., G.D., A.L-S., P.G., S.Sa.,  M.L., E.J., T.M., S.K., S.Su; Software, C.-F.G., G.D., A.L-S., P.G., M.L.; Investigation, C.-F.G., G.D., A.L-S., E.J.; Resources, P.S., R.T.; Data Curation, C.-F.G., A.L-S., G.D.; Writing – Original Draft Preparation, C.-F.G., A.L-S., P.S.; Writing – Review \& Editing, C.-F.G., G.D., A.L-S., E.J., S.Sa., S.Su., S.K., T.M.; Visualization, C.-F.G., A.L-S.; Supervision, M.Le., S.Su., S.K., T.M.; Funding Acquisition, E.J.,  M.Le., S.Su., S.K., T.M.







\supplementary

S1: Incoupling simulations. \\

S2: Waveguide and MMI splitter simulations. \\

S3: GC efficiency simulation.\\


\bibliographystyle{unsrtnat}

\begin{thebibliography}{54}
\bibitem[\protect\citeauthoryear{Biercuk et al.}{Biercuk et al.}{2010}]{biercuk_ultrasensitive_2010}
Biercuk MJ et al., Ultrasensitive detection of force and displacement using trapped ions, Nature Nanotechnology 5, 646 (2010).
\url{https://doi.org/10.1038/nnano.2010.165}

\bibitem[\protect\citeauthoryear{Harlander et al.}{Harlander et al.}{2010}]{Harlander_2010}
Harlander M et al., Trapped-ion probing of light-induced charging effects on dielectrics, New Journal of Physics 12, 093035 (2010).
\url{https://doi.org/10.1088/1367-2630/12/9/093035}

\bibitem[\protect\citeauthoryear{Maunz et al.}{Maunz et al.}{2007}]{maunz2007quantum}
Maunz P et al., Quantum interference of photon pairs from two remote trapped atomic ions, Nature Physics 3, 538 (2007).
\url{https://doi.org/10.1038/nphys644}

\bibitem[\protect\citeauthoryear{Higginbottom et al.}{Higginbottom et al.}{2016}]{higginbottom2016pure}
Higginbottom DB et al., Pure single photons from a trapped atom source, New Journal of Physics 18, 093038 (2016).
\url{https://doi.org/10.1088/1367-2630/18/9/093038}

\bibitem[\protect\citeauthoryear{Hausser et al.}{Hausser et al.}{2025}]{Hausser25}
Hausser HN et al., $^{115}\mathrm{In}^{+}$--$^{172}\mathrm{Yb}^{+}$ Coulomb crystal clock with $2.5\times10^{-18}$ systematic uncertainty, Phys. Rev. Lett. 134, 023201 (2025).
\url{https://doi.org/10.1103/PhysRevLett.134.023201}

\bibitem[\protect\citeauthoryear{Lindvall et al.}{Lindvall et al.}{2025}]{Lindvall:25}
Lindvall T et al., Coordinated international comparisons between optical clocks connected via fiber and satellite links, Optica 12, 843 (2025).
\url{https://doi.org/10.1364/OPTICA.561754}

\bibitem[\protect\citeauthoryear{Monroe and Kim}{Monroe and Kim}{2013}]{Monroe13}
Monroe C, Kim J, Scaling the ion trap quantum processor, Science 339, 1164 (2013).
\url{https://doi.org/10.1126/science.1231298}

\bibitem[\protect\citeauthoryear{Moses et al.}{Moses et al.}{2023}]{Race_track23}
Moses SA et al., A race-track trapped-ion quantum processor, Phys. Rev. X 13, 041052 (2023).
\url{https://doi.org/10.1103/PhysRevX.13.041052}

\bibitem[\protect\citeauthoryear{Mehta et al.}{Mehta et al.}{2016}]{Mehta2016}
Mehta KK et al., Integrated optical addressing of an ion qubit, Nature Nanotechnology 11, 1066 (2016).
\url{https://doi.org/10.1038/nnano.2016.139}

\bibitem[\protect\citeauthoryear{Mehta et al.}{Mehta et al.}{2020}]{Mehta2020}
Mehta KK et al., Integrated optical multi-ion quantum logic, Nature 586, 533 (2020).
\url{https://doi.org/10.1038/s41586-020-2823-6}

\bibitem[\protect\citeauthoryear{Niffenegger et al.}{Niffenegger et al.}{2020}]{niffenegger2020integrated}
Niffenegger RJ et al., Integrated multi-wavelength control of an ion qubit, Nature 586, 538 (2020).
\url{https://doi.org/10.1038/s41586-020-2811-x}
\bibitem[\protect\citeauthoryear{Bruzewicz et al.}{Bruzewicz et al.}{2019}]{Bruzewicz2019}
Bruzewicz CD et al., Trapped-ion quantum computing: Progress and challenges, Applied Physics Reviews 6, 021314 (2019).
\url{https://doi.org/10.1063/1.5088164}

\bibitem[\protect\citeauthoryear{Ivory et al.}{Ivory et al.}{2021}]{Ivory21}
Ivory M et al., Integrated Optical Addressing of a Trapped Ytterbium Ion, Phys. Rev. X 11, 041033 (2021).
\url{https://doi.org/10.1103/PhysRevX.11.041033}

\bibitem[\protect\citeauthoryear{Kwon et al.}{Kwon et al.}{2024}]{Kwon2024}
Kwon J et al., Multi-site integrated optical addressing of trapped ions, Nature Communications 15, 1 (2024).
\url{https://doi.org/10.1038/s41467-024-47882-5}

\bibitem[\protect\citeauthoryear{Knollmann et al.}{Knollmann et al.}{2024}]{Knollmann2024}
Knollmann FW et al., Integrated photonic structures for photon-mediated entanglement of trapped ions, Optica Quantum 2, 230 (2024).
\url{https://doi.org/10.1364/opticaq.522128}

\bibitem[\protect\citeauthoryear{Mordini et al.}{Mordini et al.}{2025}]{Mordini25}
Mordini C et al., Multizone Trapped-Ion Qubit Control in an Integrated Photonics QCCD Device, Phys. Rev. X 15, 011040 (2025).
\url{https://doi.org/10.1103/PhysRevX.15.011040}

\bibitem[\protect\citeauthoryear{Zesar et al.}{Zesar et al.}{2024}]{Zesa2024}
Zesar A et al., Industrial Ion Trap Chips with Integrated Optics, in Witzens J, Poon J, Zimmermann L, Freude W (eds) The 25th European Conference on Integrated Optics, Springer Nature Switzerland, Cham, p. 222 (2024).
\url{https://doi.org/10.1007/978-3-031-63378-2_37}


\bibitem[\protect\citeauthoryear{Blumenthal et al.}{Blumenthal et al.}{2018}]{Blumenthal2018}
Blumenthal DJ et al., Silicon Nitride in Silicon Photonics, Proceedings of the IEEE 106, 2209 (2018).
\url{https://doi.org/10.1109/JPROC.2018.2861576}

\bibitem[\protect\citeauthoryear{Mu\~noz et al.}{Mu\~noz et al.}{2017}]{Munoz2017}
Mu\~noz P et al., Silicon Nitride Photonic Integration Platforms for Visible, Near-Infrared and Mid-Infrared Applications, Sensors 17, 2088 (2017).
\url{https://doi.org/10.3390/s17092088}

\bibitem[\protect\citeauthoryear{Sacher et al.}{Sacher et al.}{2019}]{Sacher2019}
Sacher WD et al., Visible-light silicon nitride waveguide devices and implantable neurophotonic probes on thinned 200 mm silicon wafers, Opt. Express 27, 37400 (2019).
\url{https://doi.org/10.1364/OE.27.037400}

\bibitem[\protect\citeauthoryear{Leinse et al.}{Leinse et al.}{2016}]{Leinse2020}
Leinse A, Zhang S, Heideman R, TriPleX: The versatile silicon nitride waveguide platform, in 2016 Progress in Electromagnetic Research Symposium (PIERS), p. 67 (2016).
\url{https://doi.org/10.1109/PIERS.2016.7734240}


\bibitem[\protect\citeauthoryear{Chauhan et al.}{Chauhan et al.}{2022}]{Chauhan22}
Chauhan N et al., Ultra-low loss visible light waveguides for integrated atomic, molecular, and quantum photonics, Opt. Express 30, 6960 (2022).
\url{https://doi.org/10.1364/OE.448938}

\bibitem[\protect\citeauthoryear{Smith et al.}{Smith et al.}{2023}]{Smith23}
Smith JA et al., SiN foundry platform for high performance visible light integrated photonics, Opt. Mater. Express 13, 458 (2023).
\url{https://doi.org/10.1364/OME.479871}

\bibitem[\protect\citeauthoryear{Blasco-Solvas et al.}{Blasco-Solvas et al.}{2024}]{Blasco-Solvas:24}
Blasco-Solvas M et al., Silicon Nitride Building Blocks in the Visible Range of the Spectrum, J. Lightwave Technol. 42, 6019 (2024).
\url{https://doi.org/10.1364/JLT.42.006019}

\bibitem[\protect\citeauthoryear{Ji et al.}{Ji et al.}{2024}]{10527127}
Ji X et al., Efficient mass manufacturing of high-density, ultra-low-loss Si$_3$N$_4$ photonic integrated circuits, Optica 11, 1397 (2024).
\url{https://doi.org/10.1364/optica.529673}

\bibitem[\protect\citeauthoryear{Isichenko et al.}{Isichenko et al.}{2025}]{Isichenko2025}
Isichenko A et al., Multi-laser stabilization with an atomic-disciplined photonic integrated resonator, arXiv:2509.09124 (2025).
\url{https://arxiv.org/abs/2509.09124}

\bibitem[\protect\citeauthoryear{Corsetti et al.}{Corsetti et al.}{2026}]{corsetti_integrated-photonics-based_2026}
Corsetti SM et al., Integrated-photonics-based systems for polarization-gradient cooling of trapped ions, Light: Science \& Applications 15, 57 (2026).
\url{https://doi.org/10.1038/s41377-025-02094-4}

\bibitem[\protect\citeauthoryear{Huntemann et al.}{Huntemann et al.}{2012}]{Huntemann2012}
Huntemann N et al., High-accuracy optical clock based on the octupole transition in $^{171}\mathrm{Yb}^{+}$, Phys. Rev. Lett. 108, 090801 (2012).
\url{https://doi.org/10.1103/PhysRevLett.108.090801}

\bibitem[\protect\citeauthoryear{Tofful et al.}{Tofful et al.}{2024}]{Tofful_2024}
Tofful A et al., $^{171}\mathrm{Yb}^{+}$ optical clock with $2.2\times10^{-18}$ systematic uncertainty and absolute frequency measurements, Metrologia 61, 045001 (2024).
\url{https://doi.org/10.1088/1681-7575/ad53cd}

\bibitem[\protect\citeauthoryear{Olmschenk et al.}{Olmschenk et al.}{2007}]{Olmschenk2007}
Olmschenk S et al., Manipulation and detection of a trapped Yb$^{+}$ hyperfine qubit, Phys. Rev. A 76, 052314 (2007).
\url{https://doi.org/10.1103/PhysRevA.76.052314}

\bibitem[\protect\citeauthoryear{Piltz et al.}{Piltz et al.}{2014}]{piltz_trapped-ion-based_2014}
Piltz C et al., A trapped-ion-based quantum byte with $10^{-5}$ next-neighbour cross-talk, Nature Communications 5, 4679 (2014).
\url{https://doi.org/10.1038/ncomms5679}

\bibitem[\protect\citeauthoryear{Dreissen et al.}{Dreissen et al.}{2022}]{dreissen_improved_2022}
Dreissen LS et al., Improved bounds on Lorentz violation from composite pulse Ramsey spectroscopy in a trapped ion, Nature Communications 13, 7314 (2022).
\url{https://doi.org/10.1038/s41467-022-34818-0}

\bibitem[\protect\citeauthoryear{Hur et al.}{Hur et al.}{2022}]{Hur22}
Hur J et al., Evidence of two-source King plot nonlinearity in spectroscopic search for new boson, Phys. Rev. Lett. 128, 163201 (2022).
\url{https://doi.org/10.1103/PhysRevLett.128.163201}

\bibitem[\protect\citeauthoryear{Filzinger et al.}{Filzinger et al.}{2025}]{Filzinger25}
Filzinger M et al., Ultralight dark matter search with space-time separated atomic clocks and cavities, Phys. Rev. Lett. 134, 031001 (2025).
\url{https://doi.org/10.1103/PhysRevLett.134.031001}

\bibitem[\protect\citeauthoryear{Door et al.}{Door et al.}{2025}]{Door25}
Door M et al., Probing new bosons and nuclear structure with ytterbium isotope shifts, Phys. Rev. Lett. 134, 063002 (2025).
\url{https://doi.org/10.1103/PhysRevLett.134.063002}

\bibitem[\protect\citeauthoryear{Du et al.}{Du et al.}{2025}]{du2025electricfielddistortionssurface}
Du G, Jordan E, Mehlst\"aubler TE, Electric field distortions in surface ion traps with integrated nanophotonics, arXiv:2503.20387 (2025).
\url{https://arxiv.org/abs/2503.20387}

\bibitem[\protect\citeauthoryear{Sun et al.}{Sun et al.}{2009}]{Sun:09}
Sun X, Liu H-C, Yariv A, Adiabaticity criterion and the shortest adiabatic mode transformer in a coupled-waveguide system, Opt. Lett. 34, 280 (2009).
\url{https://doi.org/10.1364/OL.34.000280}

\bibitem[\protect\citeauthoryear{Lin et al.}{Lin et al.}{2021}]{Lin2021_BiLayerSiN}
Lin Y et al., Low-loss broadband bi-layer edge couplers for visible light, Opt. Express 29, 34565 (2021).
\url{https://doi.org/10.1364/OE.435669}

\bibitem[\protect\citeauthoryear{Zhang et al.}{Zhang et al.}{2021}]{Zhang2021_Buried3DSSC}
Zhang W et al., Buried 3D silicon photonics spot-size convertors, in 2021 IEEE 17th International Conference on Group IV Photonics (GFP), p. 1 (2021).
\url{https://doi.org/10.1109/GFP51802.2021.9673849}


\bibitem[\protect\citeauthoryear{Dai et al.}{Dai et al.}{2006}]{Dai2006_BilevelConverter}
Dai D, He S, Tsang H-K, Bilevel mode converter between a silicon nanowire waveguide and a larger waveguide, Journal of Lightwave Technology 24, 2428 (2006).
\url{https://doi.org/10.1109/JLT.2006.874554}

\bibitem[\protect\citeauthoryear{Jia et al.}{Jia et al.}{2018}]{Jia2018_SuspendedCoupler}
Jia L et al., Efficient suspended coupler with loss less than 1.4 dB between Si-photonic waveguide and cleaved single mode fiber, Journal of Lightwave Technology 36, 239 (2018).
\url{https://doi.org/10.1109/JLT.2017.2779863}

\bibitem[\protect\citeauthoryear{Yi et al.}{Yi et al.}{2024}]{Yi2024_BilevelDualCore}
Yi X et al., Asymmetric bi-level dual-core mode converter for high-efficiency and polarization-insensitive O-band fiber-chip edge coupling: breaking the critical size limitation, Nanophotonics 13, 4149 (2024).
\url{https://doi.org/10.1515/nanoph-2024-0320}

\bibitem[\protect\citeauthoryear{Chen et al.}{Chen et al.}{2025}]{Chen2025_DualSiOxN}
Chen Z et al., Modeling dual-SiOxN thin-film edge coupler with ultra-low loss and large alignment tolerance, Photonics 12, 136 (2025).
\url{https://doi.org/10.3390/photonics12020136}

\bibitem[\protect\citeauthoryear{He et al.}{He et al.}{2020}]{He2020_JLT_EdgeCouplers}
He A et al., Low loss, large bandwidth fiber-chip edge couplers based on silicon-on-insulator platform, Journal of Lightwave Technology 38, 4780 (2020).
\url{https://doi.org/10.1109/JLT.2020.2995544}

\bibitem[\protect\citeauthoryear{Marchetti et al.}{Marchetti et al.}{2019}]{Marchetti2019_CouplingReview}
Marchetti R et al., Coupling strategies for silicon photonics integrated chips, Photon. Res. 7, 201 (2019).
\url{https://doi.org/10.1364/PRJ.7.000201}

\bibitem[\protect\citeauthoryear{Mu et al.}{Mu et al.}{2019}]{Mu19}
Mu J et al., Monolithic integration of Al$_2$O$_3$ and Si$_3$N$_4$ toward double-layer active--passive platform, IEEE Journal of Selected Topics in Quantum Electronics 25, 1 (2019).
\url{https://doi.org/10.1109/JSTQE.2019.2908559}

\bibitem[\protect\citeauthoryear{Sakai et al.}{Sakai et al.}{2001}]{Sakai_bendshift}
Sakai A, Go H, Baba T, Sharply bent optical waveguide silicon-on-insulator substrate, in Physics and Simulation of Optoelectronic Devices IX (SPIE) 4283, pp. 610--618 (2001).
\url{https://doi.org/10.1117/12.432614}

\bibitem[\protect\citeauthoryear{Soldano and Pennings}{Soldano and Pennings}{1995}]{Soldano_MMI}
Soldano LB, Pennings ECM, Optical multi-mode interference devices based on self-imaging: principles and applications, Journal of Lightwave Technology 13, 615 (1995).
\url{https://doi.org/10.1109/50.372474}

\bibitem[\protect\citeauthoryear{Taillaert et al.}{Taillaert et al.}{2004}]{Taillaert:04}
Taillaert D, Bienstman P, Baets R, Compact efficient broadband grating coupler for silicon-on-insulator waveguides, Opt. Lett. 29, 2749 (2004).
\url{https://doi.org/10.1364/OL.29.002749}

\bibitem[\protect\citeauthoryear{Beck et al.}{Beck et al.}{2024}]{Gill2024}
Beck GJ, Home JP, Mehta KK, Grating design methodology for tailored free-space beam-forming, Journal of Lightwave Technology 42, 4939 (2024).
\url{https://doi.org/10.1109/JLT.2024.3381785}

\bibitem[\protect\citeauthoryear{Oton}{Oton}{2016}]{2016Oton}
Oton CJ, Long-working-distance grating coupler for integrated optical devices, IEEE Photonics Journal 8, 1 (2016).
\url{https://doi.org/10.1109/JPHOT.2015.2511098}

\bibitem[\protect\citeauthoryear{Dietl et al.}{Dietl et al.}{2025}]{Dietl25}
Dietl M et al., Test and characterization of multilayer ion traps on fused silica, Advanced Quantum Technologies 8, e00412 (2025).
\url{https://doi.org/10.1002/qute.202500412}

\bibitem[\protect\citeauthoryear{Hattori et al.}{Hattori et al.}{2024}]{Hattori:24}
Hattori A et al., Integrated visible-light polarization rotators and splitters for atomic quantum systems, Opt. Lett. 49, 1794 (2024).
\url{https://doi.org/10.1364/OL.509747}

\bibitem[\protect\citeauthoryear{Gallacher et al.}{Gallacher et al.}{2022}]{Gallacher22}
Gallacher K et al., Silicon nitride waveguide polarization rotator and polarization beam splitter for chip-scale atomic systems, APL Photonics 7, 046101 (2022).
\url{https://doi.org/10.1063/5.0077738}

\bibitem[\protect\citeauthoryear{Momenzadeh et al.}{Momenzadeh et al.}{2026}]{Momenzadeh2025}
Momenzadeh M et al., Individual trapped-ion addressing with adjoint-optimized multimode photonic circuits, npj Nanophoton. 3, 3 (2026).
\url{https://doi.org/10.1038/s44310-025-00102-4}




\end{thebibliography}

\end{document}


\title{ Benchmarking Dual-Polarization Silicon Nitride Photonic Integrated Circuits for Trapped-Ion Quantum Technologies; supplement}

\author{\firstname{Carl-Frederik} \lastname{Grimpe}\inst{1}
\and                          
\firstname{Anastasiia} \lastname{Lüßmann-Sorokina}\inst{1,2,3}\and
\firstname{Guochun} \lastname{Du}\inst{1} \fnsep\thanks{\email{guochun.du@ptb.de}} 
\and
\firstname{Pragya} \lastname{Sah}\inst{4,5}
\and
        \firstname{Steffen} \lastname{Sauer}\inst{1,2,3} \and
        \firstname{Elena} \lastname{Jordan}\inst{1} 
         \and
        \firstname{Rijil} \lastname{Thomas}\inst{4}
        \and
        \firstname{Pascal} \lastname{Gehrmann}\inst{2,3}
        \and
        \firstname{Maksim} 
        \lastname{Lipkin}\inst{4,5}  \and
        \firstname{Stephan} \lastname{Suckow}\inst{4}
        \and
        \firstname{Max C.} 
        \lastname{Lemme}\inst{4,5} 
        \and
        \firstname{Stefanie} \lastname{Kroker}\inst{1,2,3}
         \and
         \firstname{Tanja E.} \lastname{Mehlstäubler}\inst{1,6,7}
        }

\institute{Physikalisch-Technische Bundesanstalt, Bundesallee 100, Braunschweig, 38116, Germany 
\and
        Technische Universität Braunschweig, Institute of Semiconductor Technology, Hans-Sommer-Str. 66, Braunschweig, 38106, Germany 
\and
           Laboratory for Emerging Nanometrology (LENA), Langer Kamp 6a/b, Braunschweig, 38106, Germany 
\and
            AMO GmbH, Otto-Blumenthal-Straße 25, 52074 Aachen, Germany
\and
        Chair of Electronic Devices, RWTH Aachen University, Otto-Blumenthal Str. 25, Aachen 52074, Germany
\and
           Leibniz Universität Hannover, Institut für Quantenoptik, Welfengarten 1, Hannover, 30167, Germany 
\and
           Leibniz Universität Hannover, Laboratorium für Nano- und Quantenengineering, Welfengarten 1, Hannover, 30167, Germany
           }

\abstract{This Supplementary Material provides additional simulation results referenced in the main text. Section \ref{S1} summarizes the steps used in the multilayer-coupling simulations. Section \ref{S2} presents the single-mode waveguide analysis, material parameters, and multi-mode interference (MMI) splitter simulations. Section \ref{S3} includes a comparison of the simulated 2D grating coupler (GC) outcoupling efficiency for transverse-electric (TE) and transverse-magnetic (TM) modes.}

\maketitle
\newpage

\setcounter{section}{0}
\renewcommand{\thesection}{S\arabic{section}}

\section{Incoupling simulations}
\label{S1}
\begin{figure}[h]
\centering
\includegraphics[width=0.8\textwidth,clip]{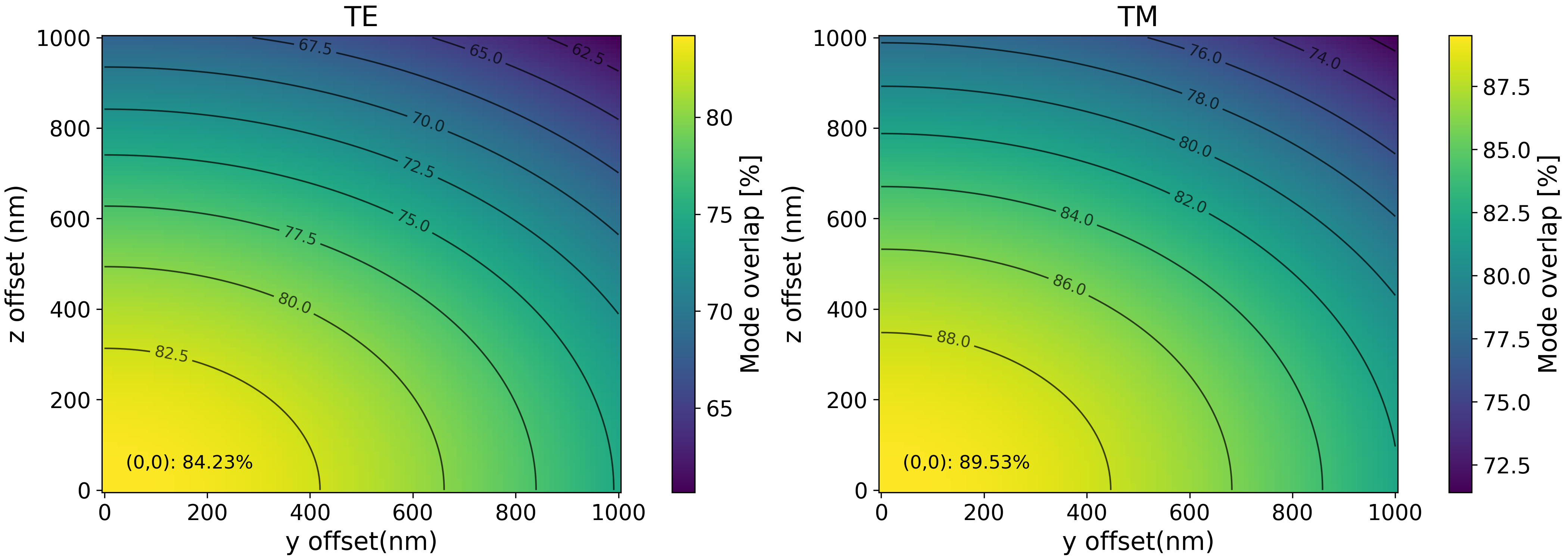}
\caption{
Simulated overlap integral between the fiber mode (core diameter 4.4\,$\upmu$m) and the WG mode (6.5\,$\upmu$m WG width, 15\,nm WG height) at 760\,nm versus transverse alignment offsets of the fiber in y- and z-direction. Results for (a) TE and (b) TM polarization.
}

\label{Modematch_simulation}
\end{figure}

\begin{figure}[h]
\centering
\includegraphics[width=0.85\textwidth,clip]{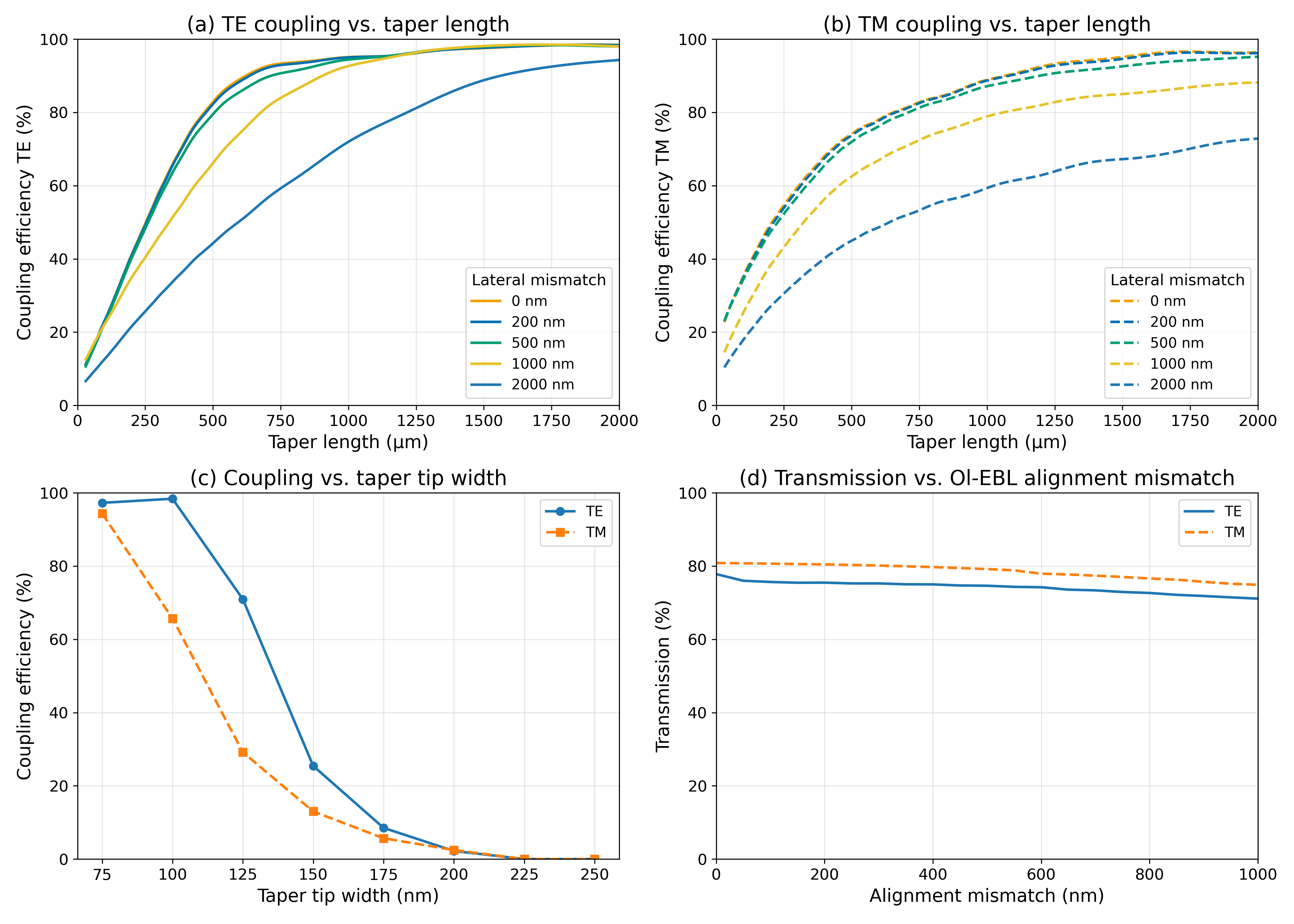}
\caption{
Two-layer coupling efficiency versus the taper length for different lateral alignment offsets for TE (a) and TM (b) modes and for different taper-tip widths of the thick-core layer (c). (d) OL-EBL taper transmission for different lateral taper alignment offsets.
}
\label{dual_layer}
\end{figure}

Fiber-to-chip edge-coupling is validated by a mode-overlap simulation using Finite-Difference Eigenmode (FDE), comparing the fundamental fiber mode with the guided mode of the photonic waveguide (see Fig.~\ref{Modematch_simulation}). Two-layer coupling was evaluated using eigenmode expansion (EME) methods, where the coupling efficiency between the waveguides is studied as a function of lateral misalignment along the taper region. The results show robust TE coupling for taper lengths exceeding 1000\,$\upmu$m, with tolerances up to approximately 1000\,nm lateral offset (see Fig.~\ref{dual_layer} a)). The TM mode exhibits a less tolerant behavior under the same conditions (see Fig.~\ref{dual_layer} b)). Fig.~\ref{dual_layer} c) shows the dependence of the taper-tip width of the thick-core Si\(_3\)N\(_4\) layer on the two-layer coupling. The influence of a lateral misalignment on the optical lithography (OL) to electron-beam lithography (EBL) taper transmission is summarized in Fig.~\ref{dual_layer} d), comparing a 100\,$\upmu$m EBL section with a 100\,$\upmu$m OL section. The design parameters and simulated coupling efficiencies are summarized in Tab. \ref{tab:coupling-efficiency}.

\begin{table*}[!hb]
\centering
\caption{Simulated coupling efficiencies (CE) for TE- and TM-polarized light at different coupling stages of the incoupling, with perfect lateral alignment (0\,nm) or a 500\,nm offset. The design parameters summarize the relevant geometrical values used in the simulations.}
\label{tab:coupling-efficiency}

\begin{tabularx}{\textwidth}{@{} l X c c c @{}}
\toprule
\makecell[l]{\textbf{Coupling stage}} &
\makecell[l]{\textbf{Design parameters}} &
\makecell{\textbf{Offset, nm}} &
\makecell{\textbf{TE CE}} &
\makecell{\textbf{TM CE}} \\
\midrule

\makecell[l]{\textbf{I: Fiber-to-chip edge}\\\textbf{coupling}} &
\makecell[l]{L-4: Si\textsubscript{3}N\textsubscript{4} width vs height\\
W: 6.5~\(\upmu\)m,\ H: 15~nm} &
\makecell{0\\500 } &
\makecell{$-0.75$ dB\\$-0.97$ dB} &
\makecell{$-0.48$ dB\\$-0.63$ dB} \\
\midrule

\makecell[l]{\textbf{II: Two-layer}\\\textbf{adiabatic coupling}} &
\makecell[l]{L-2: taper-tip width: 75~nm\\
L-4: taper-tip width: 500~nm\\
Taper length: 1400~\(\upmu\)m} &
\makecell{0\\500} &
\makecell{$-0.11$ dB\\$-0.11$ dB} &
\makecell{$-0.25$ dB\\$-0.37$ dB} \\
\midrule

\makecell[l]{\textbf{III: EBL--OL coupling}} &
\makecell[l]{L-2: EBL taper\\
W: 10~\(\upmu\)m,\ L: 100~\(\upmu\)m\\
L-2: OL taper\\
W: 10~\(\upmu\)m,\ L: 100~\(\upmu\)m} &
\makecell{0\\500} &
\makecell{$-1.09$ dB\\$-1.29$ dB} &
\makecell{$-0.92$ dB\\$-1.03$ dB} \\
\midrule

\makecell[l]{\textbf{Simulated total}\\\textbf{efficiency}} &  &
\makecell{0\\500} &
\makecell{$-1.95$ dB\\$-2.26$ dB} &
\makecell{$-1.65$ dB\\$-2.03$ dB} \\
\bottomrule
\end{tabularx}
\end{table*}

\clearpage
\newpage
\section{Waveguide and MMI splitter simulations}
\label{S2}

\begin{figure*}[h]
\centering
\includegraphics[width=0.8\textwidth,clip]{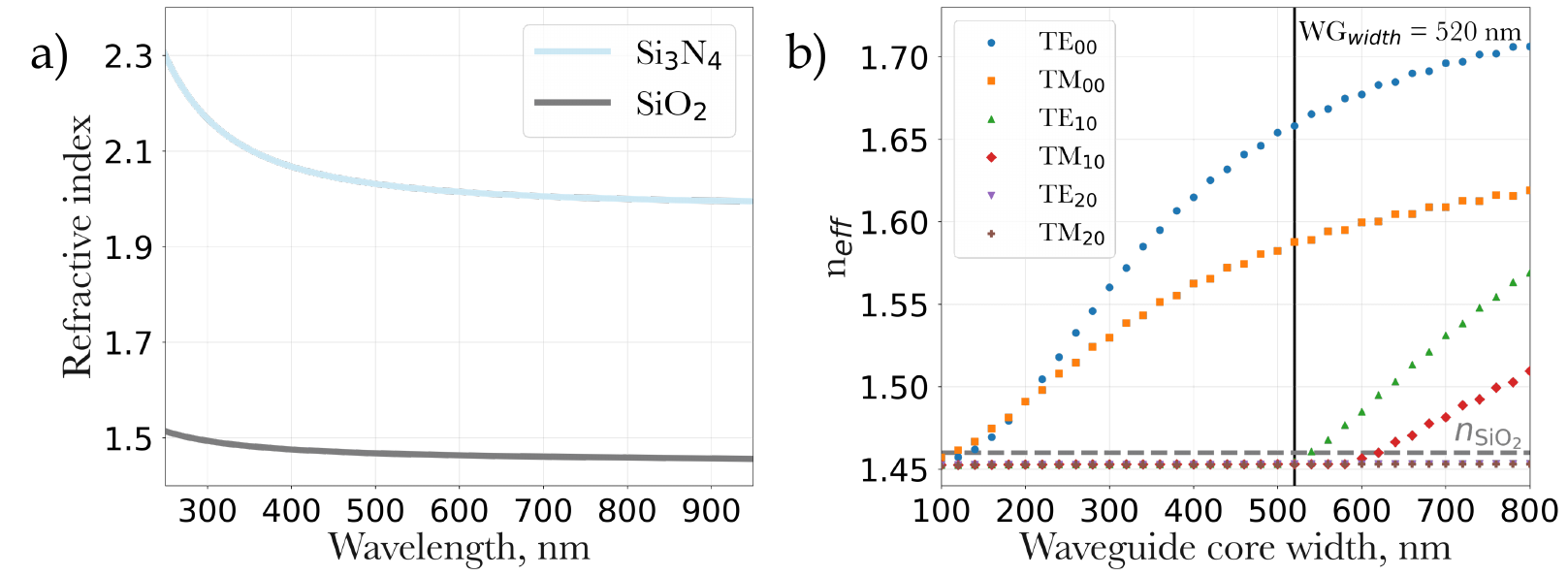}
\caption{
a) Refractive indices of Si0\(_2\) and Si\(_3\)N\(_4\) measured via ellipsometry. These values are used in all simulations presented in this work. b) Simulated effective indices of the waveguide modes as a function of core width for a 200 nm-thick Si\(_3\)N\(_4\) layer at 760 nm. The black vertical line at 520 nm indicates the waveguide width selected for the final design, ensuring single-mode operation.
}
\label{wg_sm}
\end{figure*}

The refractive indices used in the simulations are obtained from ellipsometry measurements (see Fig.~\ref{wg_sm} a)). FDE simulations are then performed to determine the maximum waveguide width that exhibits single-mode behavior, defined as the width at which the effective index of the higher-order modes exceeds the refractive index of the silicon dioxide (SiO\textsubscript{2}) cladding.

\begin{figure*}[!ht]
\centering
\includegraphics[width=\textwidth,clip]{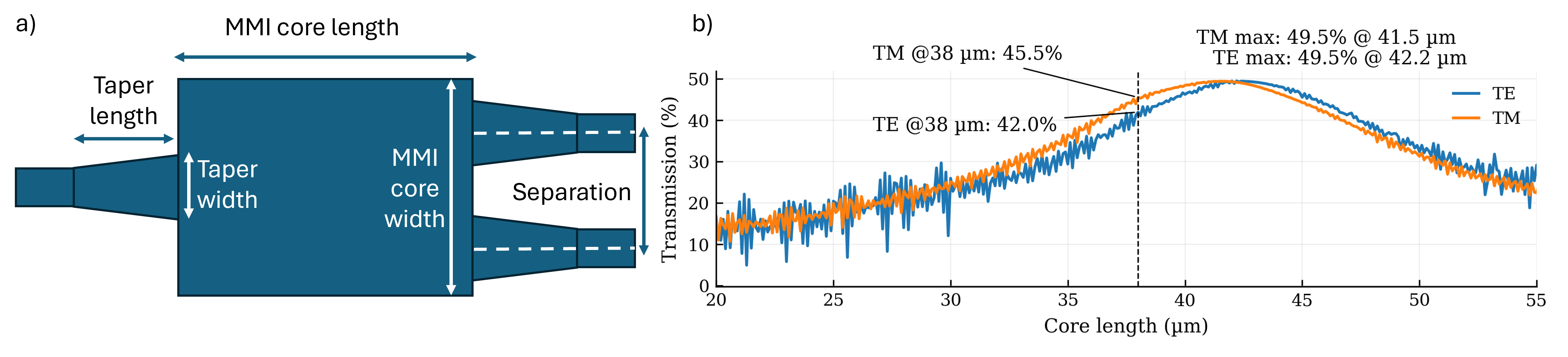}
\caption{
a) Schematic of a MMI splitter and design parameters. MMI core length of 38\,$\upmu$m, core width of 6\,$\upmu$m, separation of 3.14\,$\upmu$m and taper width of 1.6\,$\upmu$m. b) Simulated TE and TM transmission as a function of MMI core length. The plot highlights the maximum transmission efficiencies for both polarizations, as well as the efficiencies at the design core length of 38 \,$\upmu$m. 
}
\label{MMI_simulation}
\end{figure*}

Figure~\ref{MMI_simulation} a) shows the schematic of the MMI splitter explaining the parameters used in the simulations. Figure~\ref{MMI_simulation} b) presents a sweep of the core length for TE and TM polarization and the corresponding transmission at each output port. The optimal length for maximal efficiency, along with the chosen design length, is indicated in the plot.

\newpage

\section{GC efficency simulations}
\label{S3}

GC efficiencies for the TE and TM modes are compared using two-dimensional finite-difference time-domain (FDTD) simulations. The outcoupling efficiency is overall lower for the TM mode than for the TE mode. Outcoupling efficiencies of 40\% for the TM mode can only be achieved with long gratings of 300~$\upmu$m or longer, whereas the same efficiency can be reached with a grating length of 50~$\upmu$m for the TE mode. The duty cycle (DC) was fixed at 0.5, and the grating period was calculated using the Bragg condition to target a $-30^{\circ}$ emission angle. All simulations were performed in SiO\textsubscript{2}.

\begin{figure*}[!ht]
\centering
\includegraphics[width=\textwidth,clip]{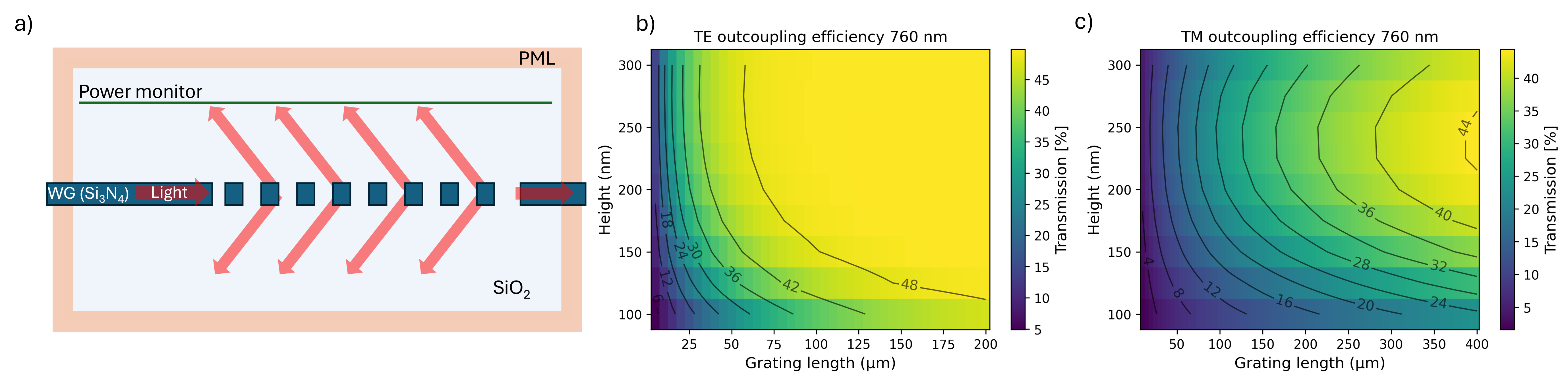}
\caption{
a) Schematic of the 2D simulation region. The period was calculated via the Bragg condition. The DC was set to 0.5. Target angle in Si0\(_2\) was set to -30\,$^{\circ}$. b) and c) simulated outcoupling efficiencies (upward diffracted light) as a function of GC length and layer height for TE and TM modes, respectively. }
\label{GC_simulation_eff}
\end{figure*}